\documentclass[aps,prl,preprintnumbers,epsf]{revtex4}
\newcommand{\gtap}{\;{\raise.3ex\hbox{$>$\kern-.75em\lower1ex\hbox{$\sim$}}}\;}
\newcommand{\ltap}{\;{\raise.3ex\hbox{$<$\kern-.75em\lower1ex\hbox{$\sim$}}}\;}
\usepackage[dvips]{graphicx}
\newcommand{\bea}{\begin{eqnarray}}
\newcommand{\eea}{\end{eqnarray}}
\begin{document}
\preprint{hep-th/0507024v3}
\title{Renormalization group equations as 'decoupling' theorems}
\author{Ji-Feng Yang}
\address{II. Institut f\"ur Theoretische Physik, Universit\"at
Hamburg, 22761 Hamburg, Germany\\
Department of Physics, East China Normal University, Shanghai
200062, China\footnote{Permanent address.}}
\begin{abstract}We propose a simple derivation of renormalization
group equations and Callan-Symanzik equations as decoupling
theorems of the structures underlying effective field theories.
\end{abstract}\maketitle
\section{Introduction}
It is well known that any known quantum field theory could at best
be an effective field theory (EFT) that include ingredients as
many as possible for the description of certain phenomena. Various
divergences (UV, IR and other types) arise in an EFT simply due to
its simplification made at the scales widely separated from the
dominating ones. In other words, an EFT is usually not a complete
framework for accounting for the quantum fluctuations over any
distances, it fails for certain modes underlying the 'effective
ones'.

The conventional ways to deal with the divergences within EFT
framework are renormalization (for UV divergences) and
factorization (for IR and/or collinear divergences): The true
non-calculable contributions from the underlying structures could
be separated and put into some EFT ingredients (operators,
couplings or matrix elements, etc.) that could be determined
later. Such tricky procedures lead to the celebrated
renormalization group equations (RGE). However, in the
conventional treatment, an artificial regularization (as simple
and temporary substitutes for the true underlying structures) must
be employed in the course of calculations and the associated
tricky procedures seem to obscure the physical rationality behind
the treatments, though the final results should be independent of
such regularization procedures.

In this presentation, I wish to give a rather simple (to some
extent, less rigorous) derivation of RGE and the Callan-Symanzik
equation, solely based on the simple picture that there are some
elaborate details underlying the effective theories in use for
certain purpose. In other words, the scale hierarchy
substantiating the EFT construction is the only starting point. It
is hoped that such simple lines of arguments could help the
further exploration of the more fundamental implications of the
EFT reconstructions as well as various factorization approaches
that prevail in physics literature.
\section{Effective versus underlying structures}
To proceed, we take the existence of a well-defined formulation of
the theory underlying (UT) the EFT's as a natural fact or
postulate. In fact, we could take the well defined formulation of
an EFT as a result of reconstruction or projection procedure from
the related sectors in the complete UT. The hierarchy between the
scales in the formulation discriminates the parameters into
effective and underlying ones, respectively. Such hierarchy should
automatically facilitate a well defined expansion in terms of the
ratios like $\Lambda_{EFT}/\Lambda_{UT}$ (for UV underlying
structures) or $\Lambda_{UT}/\Lambda_{EFT}$ (for IR or other
non-UV underlying structures), so that the resulting formulation
is expressed in terms of the effective parameters and some
possible 'agent' constants (only arise in the loop contributions)
in the limit that the underlying details are apparently
'decoupled'. No unphysical divergences should appear in the course
of such reconstruction or projection procedure.

For convenience, we introduce the symbol $\breve{P}_{\texttt{\tiny
EFT}}$ to assume all the elaborate procedures of the
reconstruction or projection of an EFT out of a complete
underlying theory, which should contain at least the following
three operations: (1) projecting into the subspace of EFT; (2)
averaging over the associated underlying dynamical processes
(integrating out); (3) taking the decoupling limits with respect
to the typical underlying constants, $\{\sigma\}$. With the help
of this projection symbol, we could easily identify the technical
origin of various divergences in EFT. To see this point, we employ
the path integral formalism.

According to the above arguments, a well defined generating
functional for an EFT should be obtained from the EFT projection
on the generating functional for UT, $\breve{P}_{\texttt{\tiny
EFT}}$. \bea\label{EFTprojection} Z_{\texttt{\tiny
EFT}}([J_{\texttt{\tiny EFT}}])\equiv{\breve{P}}_{\texttt{\tiny
EFT}}\int d[\Phi] \exp\{i
{\mathcal{S}}([\Phi],[g];\{\sigma\}\|[J_{\texttt{\tiny
EFT}}])\},\eea with $[g]$ being the 'effective' parameters. That
means, the path integral should be performed in the presence of
the underlying structures. If one perform the projection first
before the path integral, an ill-defined EFT is resulted,\bea \int
d[\phi] \exp\{ i
{\mathcal{S}}([\phi],[g])\}_{\texttt{ill-defined}}= \int
 {\breve{P}}_{\texttt{\tiny EFT}} d[\Phi] \exp\{ i
{\mathcal{S}}([\Phi],[g];\{\sigma\})\}.\eea Thus, the appearance
of various divergences in an EFT implies that path integral and
the EFT projection ${\breve{P}}_{\texttt{\tiny EFT}}$ do not
commute.

In conventional approaches, the role of the sophisticated
projection procedure $ {\breve{P}}_{\texttt{\tiny EFT}}$ is played
by the ${\mathcal{R}}$ operation procedure in the well known BPZH
program when there is only UV divergences. That is,
${\mathcal{R}}$ operation could be seen as one tricky realization
of the projection operation ${\breve{P}}_{\texttt{\tiny EFT}}$ (in
the case of UV divergences alone), which could in turn be put into
the forms with multiplicative renormalization,\bea
{\breve{P}}_{\texttt{\tiny EFT}}(\hat {O})\Rightarrow
{\mathcal{R}}(\hat{O})\Rightarrow Z_O^{-1}\hat{O}.\eea
\section{Canonical scaling with
underlying structures} Now let us consider a general vertex
function (1PI) $\Gamma^{(n)}([p],[g];\{\sigma\})$ that is well
defined in UT with $[p],[g]$ denoting the external momenta and the
Lagrangian couplings (including masses) in an EFT and $\{\sigma\}$
denoting the underlying parameters or constants. Now it is easy to
see that such a vertex function must be a homogeneous function of
all its dimensional arguments, that is
\begin{eqnarray}
\label{scaling1} \Gamma^{(n)}([\lambda p],[\lambda^{d_g} g]; \{
\lambda^{d_{\sigma}} \sigma\})= \lambda^{d_{\Gamma^{(n)}}}
\Gamma^{(n)} ([p],[g];\{\sigma\})
\end{eqnarray} where $d_{\cdots}$ refers to the canonical mass
dimension of any parameters involved.

The corresponding equation in EFT could be obtained through the
application of the projection $\breve{P}_{\texttt{\tiny EFT}}$ to
both sides of Eq.(\ref{scaling1}),
\begin{eqnarray}
\label{scaling2} \Gamma ^{(n)}([\lambda p],[\lambda
^{d_g}g];\{\lambda ^{d_{\bar{c}}}{\bar{c}}\})=\lambda ^{d_{\Gamma
^{(n)}}}\Gamma ^{(n)}([p],[g];\{{\bar{c}}\}).
\end{eqnarray} Here some constants $\{\bar{c}\}$ must
appear as the agents of $\{\sigma\}$ to maintain the dimension
balance between $\{\sigma\}$ and the EFT couplings and masses.
Note that $\{{\bar{c}}\}$ only appear in the loop diagrams of EFT.

The differential form for Eq.(\ref{scaling1}) reads
\begin{eqnarray}
\label{scaling3} \{\lambda \partial _\lambda  + \sum d_g
g\partial_g + \sum d_{\sigma} \sigma \partial_{\sigma}
-d_{\Gamma^{(n)}}\} \Gamma^{(n)}([\lambda p],[g];\{\sigma\})=0.
\end{eqnarray}Since $\sum d_g g\partial_g=-i\Theta$ (the trace of
the stress tensor), the alternative form of Eq.(\ref{scaling3})
reads,\begin{eqnarray} \label{scaling4} \{\lambda
\partial _\lambda  + \sum d_{\sigma} \sigma
\partial_{\sigma} -d_{\Gamma^{(n)}}\}
\Gamma^{(n)}([p],[g];\{\sigma\}) =i\Gamma^{(n)}_{\Theta}
([0;\lambda p],[g];\{\sigma\}).\end{eqnarray} Obviously only
dimensional constants contribute to the scaling behavior.
Eq.(\ref{scaling3}) or Eq.(\ref{scaling4}) is just the most
general UT version of the EFT scaling laws. They differ from naive
EFT scaling laws only by the {\bf canonical} contribution from the
underlying structures ($\sum d_{\sigma} \sigma
\partial_{\sigma}$). This is just the origin of EFT scaling
anomalies.

To see this point, we first note the consequence of applying
$\breve{P}_{\texttt{\tiny EFT}}$ to $\sum d_{\sigma} \sigma
\partial_{\sigma}$,
\begin{eqnarray} \label{decoupling} \breve{P}_{\texttt{\tiny EFT}}\{\sum
d_{\sigma}\sigma
\partial_{\sigma}\} \Gamma ^{(n)}([\cdots];\{\sigma \})=
 \{ \sum d_{\bar c} {\bar c} \partial_{\bar c}  \}
\Gamma^{(n)}([\cdots];\{\bar c\}).
\end{eqnarray}Then, it is straightforward
to see that, within EFT, $\sum d_{\bar c}{\bar c}\partial_{\bar
c}$ has to be expanded into the insertion of appropriate EFT
operators ($[I_{O_i}]$, 'elementary' or composite) with
appropriate coefficients ($\delta_{O_i}$). Thus, we arrive at the
following decoupling theorem,
\begin{eqnarray}
\label{preRGE}
 &&\breve{P}_{\texttt{\tiny EFT}}\{\sum
d_{\sigma}\sigma
\partial_{\sigma}\}=\sum d_{\bar{c}}{\bar{c}}\partial
_{\bar{c}}=\sum_{O_i} \delta_{O_i}I_{O_i}.
\end{eqnarray} Note that each $\delta_{O_i}$ must at least be a
function of EFT couplings $[g]$ and $\{\bar c\}$. At present
stage, we do not exclude nonlocal operators from the set of
$[O_i]$, which might be more relevant in the presence of IR or
other non-UV divergences. Thus Eq.(\ref{preRGE}) is a rather
general form of 'decoupling theorem' for any sort of underlying
structures, UV or IR.

As the final step, it is easy to classify these operators into
the kinetic operators (for the EFT fields $[\phi]$), the coupling
operators (with couplings $[g]$), and 'composite' ones, $[O_N]$,
that do not appear in the EFT lagrangian,
\begin{eqnarray}
\label{decomposition} &&\sum_{O_i} \delta_{O_i}
I_{O_i}=\sum_{g}\delta _g g\partial _g+\sum_{\phi}\delta _\phi
{\hat{I}}_\phi +\sum_{O_N}\delta _{O_N}{\hat{I}}_{O_N}.
\end{eqnarray}
Now with Eqs.(\ref{decoupling},\ref{decomposition}) we can turn
the primary decoupling theorem in Eq.(\ref{preRGE}) and the full
scaling law in Eq.(\ref{scaling3}) into the following forms,
\begin{eqnarray}
\label{preRGE1}&&\{\sum_{\bar c} d_{\bar{c}}{\bar{c}}\partial
_{\bar{c}}-\sum_{O_N}\delta _{O_N}{\hat{I}} _{O_N}-\sum_{g}\delta
_gg\partial_g -\sum_{\phi}\delta _\phi { \hat{I}}_\phi\}
\Gamma ^{(n)}([\lambda p],[g];\{\bar{c}\})=0,\\
\label{scaling11} &&\{\lambda \partial _\lambda +\sum_{O_N}\delta
_{O_N}{\hat{I}} _{O_N}+\sum_{g}(d_g+\delta _g)g\partial
_g+\sum_{\phi}\delta _\phi { \hat{I}}_\phi -d_{\Gamma
^{(n)}}\}\Gamma ^{(n)}([\lambda p],[g];\{\bar{c}\})=0.
\end{eqnarray}Here Eqs.(\ref{preRGE1},\ref{scaling11}) are only
true for the complete sum of all graphs (or up to a certain
order). It is obvious that the {\bf canonical} contributions from
the underlying structures become the {\bf anomalies} in terms of
EFT parameters. $\delta_{O_i}$ is just the anomalous dimension for
the operator $O_i$ in EFT. Again the operators contributing to the
scaling anomalies should contain the ones corresponding to IR or
other non-UV singularities.
\section{Novel perspective of RGE and Callan-Symanzik Equation}
In this section we limit our attention to a special type of
theories beset only with UV divergences: the ones without the
scaling anomalies $\sum_{\{O_N\}} \delta_{O_N}([g];\{\bar c\}) {
\hat{I}}_{O_N}$, i.e., the renormalizable theories in conventional
terminology, and all the operators are now local. Then, Eqs.
(\ref{preRGE1}) and (\ref{scaling11}) become simpler,
\begin{eqnarray}
\label{preRGE2} &&\{\sum_{\bar{c}}d_{\bar{c}}\bar{c}\partial
_{\bar{c}}-\sum_{g}\delta _g g\partial _g-\sum_{\phi }\delta _\phi
{\hat{I}}_\phi \}
\Gamma ^{(n)}([\lambda p],[g];\{\bar{c}\})=0; \\
\label{scaling13} &&\{\lambda \partial _\lambda +\sum_{g}
(d_g+\delta _g)g\partial _g+\sum_{\phi} \delta _\phi
{\hat{I}}_\phi -d_{\Gamma ^{(n)}}\}\Gamma ^{(n)}([\lambda
p],[g];\{\bar{c}\})=0.
\end{eqnarray}
These equations just correspond to the usual RGE and
Callan-Symanzik equation (CSE) for renormalizable theories. We
could turn these equations into more familiar forms. For this
purpose, we note that all the agent constants could be
parametrized in terms of a single scale $\bar{\mu}$ and a series
dimensionless ones $({\bar c}_0)$. In the conventional programs,
they are first predetermined through renormalization conditions,
finally transformed into the physical parameters\cite{sterman} or
fixed somehow\cite{scheme}.

\subsection{RGE and CSE as decoupling theorems}
In Eq.(\ref{preRGE2}) and (\ref{scaling13}) only the insertion of
kinetic operators appears unfamiliar. To remove this
unfamiliarity, let us note that $\sum_{\phi }\delta _\phi
{\hat{I}}_\phi$ lead to the following
consequences\cite{yangscaling}:
\begin{eqnarray}
\label{scaling16} &&\delta_g \rightarrow  {\bar \delta}_g\equiv
(\delta_g- n_{g;\phi}\frac{\delta_{\phi}}{2}-n_{g;\psi}
\frac{\delta_{\psi}}{2}),\ \ \Gamma^{(n_{\phi},n_{\psi})}
\rightarrow
(1+\delta_{\psi})^{n_{\psi}/2}(1+\delta_{\phi})^{n_{\phi}/2}
\Gamma^{(n_{\phi},n_{\psi})},
\end{eqnarray}with $n_{g;\phi}$ and $n_{g;\psi}$ being
respectively the number of bosonic and fermionic field operators
contained in the vertex with coupling $g$.

Then Eq.(\ref{preRGE2}) and (\ref{scaling13}) take the following
forms:
\begin{eqnarray}
\label{RGE} &&\{{\bar{\mu}}\partial_{\bar \mu} - \sum_{g} {\bar
\delta}_g g\partial_g - \sum_{\phi}n_{\phi}
\frac{\delta_{\phi}}{2}- \sum_{\psi}n_{\psi}\frac{
\delta_{\psi}}{2} \} \Gamma^{(n_{\phi},n_{\psi})}
([p],[g];\{\bar{\mu};(\bar{c}^i_0)\})=0;\\
\label{scaling18} & &\{ \lambda \partial_{\lambda}+ \sum_{g} D_gg
\partial_g + \sum_{\phi}n_{\phi}\frac{\delta_{\phi}}{2} +
\sum_{\psi} n_{\psi}\frac{\delta_{\psi}}{2}
-d_{\Gamma^{(n_{\phi},n_{\psi})}}\}\Gamma^{(n_{\phi},n_{\psi})}
([\lambda p],[g];\{\bar{\mu};(\bar{c}^i_0)\}) =0,
\end{eqnarray} with $D_g \equiv {\bar
\delta}_g +d_g$. Eqs.(\ref{RGE},\ref{scaling18}) take the familiar
forms of RGE and CSE. Here, all the constants (finite!) survive or
arise from the 'decoupling' limit implied by the projection
procedure. Thus, we could naturally interpret these equations as
'decoupling' theorems for the canonical scaling laws with
underlying structures.

The RGE and CSE for the generating functional read,
\begin{eqnarray}
\label{RGEGen}&&\{{\bar{\mu}}
\partial_{\bar{\mu}}-\sum_{g}\delta_gg\partial_g
-\sum_{\phi} \delta_{\phi}{\hat{I}}_{\phi}\}  \Gamma^{1PI}
([\phi],[g];\{\bar{\mu};(\bar{c}^i_0)\})=0;\\
 \label{ScalingWT}&&\left \{\sum_{\phi}\int d^D x [d_{\phi}-x\cdot\partial_x)
\phi(x)] \frac {\delta}{\delta \phi(x)} + \sum_{g}D_gg\partial_g
+\sum_{\phi} \delta_{\phi}{\hat{I}}_{\phi}-D\right \}
\nonumber\\&&\times\Gamma^{1PI}
([\phi],[g];\{\bar{\mu};(\bar{c}^i_0)\}) =0.
\end{eqnarray}
with $D$ denoting the spacetime dimension. We note that the
operator trace anomalies\cite{Trace} could be readily read from
Eq.(~\ref{ScalingWT}),
\begin{eqnarray}
\label{scaling26}g_{\mu \nu }{\Theta }^{\mu \nu
}=-\sum_{g}\delta_gg\partial_g {\mathcal{L}}_{\texttt{\tiny
EFT}}+\sum_{\phi} \delta_{\phi}{\hat{O}}_{kinetic}(\phi).
\end{eqnarray} The right hand side could be further simplified
after using motion equations.

\subsection{RGE and CSE for composite operators}
For the vertex functions in the presence of composite operators,
the only complication lies in the contribution from the related
composite operators (more could show up than those in the vertex
functions), that is the general cases in
Eqs.(\ref{preRGE1},\ref{scaling11}),\bea
\label{compoRGE0}&&\{\bar{\mu}\partial
_{\bar{\mu}}-\sum_{O_N}\bar{\delta} _{O_N}{\hat{I}}
_{O_N}-\sum_{g}\bar{\delta}_gg\partial_g - \sum_{\phi}n_{\phi}
\frac{\delta_{\phi}}{2}- \sum_{\psi}n_{\psi}\frac{
\delta_{\psi}}{2} -\sum_{i=A,\cdots}\bar{\delta}_{O_i}\}\nonumber\\
&&\times
\Gamma ^{(n)}_{O_A,\cdots}([\lambda p],[g];\{\bar{\mu};(\bar{c}^i_0)\})=0,\\
\label{compoCSE0}&&\{\lambda \partial _\lambda
+\sum_{O_N}\bar{\delta }_{O_N}{\hat{I}}
_{O_N}+\sum_{g}D_gg\partial _g+ \sum_{\phi}n_{\phi}
\frac{\delta_{\phi}}{2}+ \sum_{\psi}n_{\psi}\frac{
\delta_{\psi}}{2}  -D_{\Gamma
^{(n)}_{O_A,\cdots}}\}\nonumber\\
&&\times\Gamma ^{(n)}_{O_A,\cdots}([\lambda
p],[g];\{\bar{\mu};(\bar{c}^i_0)\})=0. \eea  Here $D_{\Gamma
^{(n)}_{O_A,\cdots}}=d_{\Gamma
^{(n)}_{O_A,\cdots}}-\sum_{i=A,\cdots}\bar{\delta}_{O_i}$, while
the contributions from other composite operators
($\sum_{O_N}\bar{\delta }_{O_N}{\hat{I}} _{O_N}$) could be further
put into the familiar forms as the non-diagonal anomalous
dimensions:$\sum_{O_N}\bar{\delta }_{O_N}{\hat{I}} _{O_N}\Gamma
^{(n)}_{O_A,\cdots}=\sum_{[O_N],[i=A,\cdots]}\bar{\delta}_{O_NO_i}
\Gamma ^{(n)}_{O_A,\cdots}$. Thus, the final forms for
Eqs.(\ref{compoRGE0},\ref{compoCSE0}) read
\bea\label{compoRGE1}&&\{\bar{\mu}\partial
_{\bar{\mu}}-\sum_{g}\bar{\delta}_gg\partial_g -
\sum_{\phi}n_{\phi} \frac{\delta_{\phi}}{2}-
\sum_{\psi}n_{\psi}\frac{ \delta_{\psi}}{2}
-\sum_{[O_N],[i=A,\cdots]}\bar{\delta}_{O_NO_i}
-\sum_{O_A,\cdots}\bar{\delta}_{O_i}\}\nonumber\\
&&\times
\Gamma ^{(n)}_{O_A,\cdots}([\lambda p],[g];\{\bar{\mu};(\bar{c}^i_0)\})=0,\\
\label{compoCSE1}&&\{\lambda \partial _\lambda
+\sum_{g}D_gg\partial _g+ \sum_{\phi}n_{\phi}
\frac{\delta_{\phi}}{2}+ \sum_{\psi}n_{\psi}\frac{
\delta_{\psi}}{2} +\sum_{[O_N],[i=A,\cdots]}\bar{\delta}_{O_NO_i}
-D_{\Gamma
^{(n)}_{O_A,\cdots}}\}\nonumber\\
&&\times\Gamma ^{(n)}_{O_A,\cdots}([\lambda
p],[g];\{\bar{\mu};(\bar{c}^i_0)\})=0. \eea
\subsection{Underlying structures and the notion of renormalization}
As remarked above, the EFT parameters $[g]$ and the UT agents
$\{\bar c\}$ should be 'derived' from UT (they could not be
derived from EFT!). Thus, in EFT, they have to be determined or
fixed somehow through physical boundaries or data\cite{sterman} or
through sensible procedures\cite{scheme}.

Now to see the origin of the notion of renormalization, we solve
the equation for scaling law, Eq.(\ref{scaling18}). This could be
conveniently achieved through the introduction of 'running'
parameters $[\bar{g}(\lambda)]$ for $[g]$ based on Coleman's
bacteria analogue\cite{Coleman}. Then the solution of
Eq.(\ref{scaling18}) can be found as the solution of the following
equation,
\begin{eqnarray}
\label{bacteria} &&\{\lambda \partial _\lambda +\sum_{\bar g}
[d_{\bar g}+\delta _{\bar g}([\bar g];\{\bar c\})]{\bar g}\partial
_{\bar g}+\sum_{\phi} \delta _\phi ([\bar g];\{\bar c\})
{\hat{I}}_\phi -d_{\Gamma ^{(n)}}\}\Gamma ^{(n)}([\lambda p],[\bar
g];\{\bar{c}\})=0
\end{eqnarray}with $\bar g(={\bar
g}([g];\lambda))$ satisfying the following kind of equation,
\begin{eqnarray}
\label{running} \lambda \partial _\lambda \{{\bar
g}([g];\lambda)/\lambda^{d_g}\}=-\{d_{\bar g}+\delta _{\bar
g}([\bar g];\{\bar c\})\}{\bar g}([g];\lambda)/\lambda^{d_g},
\end{eqnarray}with the natural boundary condition: ${\bar
g}([g];\lambda)|_{\lambda=1}=g$ for each EFT parameter. The EFT
couplings $[g]$ should be finite 'bare' parameters as they are in
principle defined in the underlying theory. Now the notion
renormalization arises in EFT with the rescaling
$[p]\rightarrow[\lambda p]$: the EFT couplings $[g]$ (defined in
UT!) get 'renormalized', $[g]\rightarrow[{\bar g}([g];\lambda)]$.
Accordingly, one could define the 'renormalization' constants
(finite again!) as $z_g([g];\lambda)\equiv{\bar
g}([g];\lambda)/g$. Thus renormalization is a notion in EFT
associated with the rescaling, whose genuine origin is the
contributions from underlying structures.

The renormalization constants for operators (kinetic or composite)
could be introduced in similar manner\cite{yangscaling}, including
the cases with 'mixing'\cite{opmixing}. As a result, in the
underlying theory point of view, the 'renormalization' constants
are finite and could be introduced afterwards as byproducts, not
as compulsory components. We suspect that this simple scenario
might be helpful in more complicated EFT's, e.g., the Standard
model, especially in its sectors with unstable fields and with
flavor mixing.

\section{Appelquist-Carazzone decoupling theorem and underlying structures}
Now let us discuss the decoupling theorem a la
Appelquist-Carazzone\cite{Appel} from the underlying structures'
perspective.
\subsection{Decoupling and repartition}
First let us note that, in the underlying theory perspective, the
EFT parameters and the underlying parameters are grouped or
partitioned into two separate sets by the reference scale that
naturally appear in any physical processes (e.g., center energy in
a scattering ) according to the relative magnitudes: the effective
set $[g]$ and the underlying set $\{\sigma\}$. When an EFT field
'becomes' too heavy to directly participate the EFT dynamics, it
only induces a new partition between the effective and underlying
parameters, with the union of the two sets kept 'conserved' in the
course of decoupling:
\begin{eqnarray}
\label{repartition} && [g]\bigcup\{\sigma\}=
[g]^{\prime}\bigcup\{\sigma\}^{\prime},\ \ [g]^{\prime}\equiv
[g]/[M_H],\ \ \{\sigma\}^{\prime}\equiv \{\sigma\}\bigcup [M_H].
\end{eqnarray}This repartition yields a new EFT that differs from
the original one by a very massive field, hence a new set of agent
constants $\{{\bar c}\}^{\prime}$ is generated from this
repartition. In terms of the scaling behavior, that means,
\begin{eqnarray}
\label{rearrange} &&\sum_\sigma d_\sigma\sigma
\partial_\sigma+ \sum_gd_g g\partial_g=
\sum_{\sigma}^{\prime} d_\sigma\sigma
\partial_\sigma+ \sum_g^{\prime}d_g g\partial_g,
\end{eqnarray} or, equivalently,\begin{eqnarray}
 \label{rearrange2}&&\sum_{\bar c}d_{\bar c} \bar
{c}\partial_{\bar c} + \sum_gd_g
g\partial_g\Longrightarrow\sum_{{\bar c}^{\prime}}d_{{\bar
c}^{\prime}} {\bar c}^{\prime}\partial_{{\bar c}^{\prime}} +
\sum_g^{\prime}d_g g\partial_g.
\end{eqnarray}

Then from the configuration of the parameters described in
Eq.(\ref{repartition}), the decoupling of a heavy EFT field could
be formally be taken as the following two operations: (i)
repartitioning the EFT and underlying parameters; (ii) taking the
low energy limit with respect to the new underlying parameters,
including the heavy EFT field's mass. This is the mathematical
formulation of EFT field decoupling.\subsection{Practical
decoupling of EFT fields} From Sec.IV.C., we have seen that, the
underlying parameters or their agents scale contribute to the
'running' or 'renormalization' of the EFT parameters. Thus the
repartition will alter the contents of 'running': what to scale as
underlying parameters versus what to scale as EFT parameters:

(a). Before decoupling, namely, $M_H$ is an EFT parameter and does
not vary together with the underlying parameters or their agents.
Then the factor $\ln \frac{{\bar c}^2}{M_H^2}$ contributes to the
running of any EFT objects,
\begin{eqnarray}
&&\delta_\lambda\left [\ln \frac{{\bar c}^2}{M_H^2}\right ]=\ln
\frac{[(1+\delta \lambda){\bar c}]^2}{M_H^2}-\ln \frac{{\bar
c}^2}{M_H^2}\neq0;
\end{eqnarray}

(b). After decoupling, $M_H$ becomes underlying and varies
homogeneously with $\{\bar c\}$, then,\begin{eqnarray}
&&\delta_\lambda\left [\ln \frac{{\bar c}^2}{M_H^2}\right ]=\ln
\frac{[(1+\delta \lambda){\bar c}]^2}{[(1+\delta
\lambda)M_H]^2}-\ln \frac{{\bar c}^2}{M_H^2}=0.\end{eqnarray}That
is, taking as a member of the underlying parameters or their
agents, $M_H$ will cancel out some agents' contributions to the
'running'. Accordingly, the anomalous dimensions of the EFT
parameters will alter by a finite amount ($\Delta{\bar \delta}_g$)
due to decoupling:\begin{eqnarray}{\bar \delta}_g\Rightarrow {\bar
\delta}^{\prime}_g={\bar \delta}_g+\Delta{\bar
\delta}_g.\end{eqnarray} From Eq.(\ref{rearrange}) or
(\ref{rearrange2}), we see that, one could also work with the EFT
containing $M_H$, only adding some unnecessary technical
complexities. This is also a well-known fact.

It is not difficult to see that our arguments above amounts to
provide a relatively 'physical' rationale to the 'subtraction'
solution of decoupling\cite{EFT}. Of course different boundary
conditions are needed across the threshold, which must lead to
certain matching conditions for the 'running'
parameters\cite{matching}. We hope our understanding could also be
useful in the heavy quark effects in deeply inelastic
scattering\cite{PDF} and other important phenomenologies.
\section{Discussion and summary}
Of course, all the results presented here are not new. However, we
still feel the way we derived these results seems more general and
more natural. This underlying theory perspective might be of helps
in deepening our understanding of the EFT methods. This is
relevant to all the quantum theories, as any known theory is in
fact an effective theory to some extents.

Of course, we did not touch the cases with concrete IR or other
non-UV singularities. In such cases the underlying structures at
large distances (hence often nonperturbative) must be incorporated
(expanded) in a controllable way. Considering the complications in
various factorization formulations, we refrain here from a na\"ive
extrapolation of the scaling law. But the abstract operator form
of the decoupling theorems (or scaling laws) in Eq.(\ref{preRGE})
(or Eq.(\ref{scaling11})) is valid independent of such concrete
details. The next step for deriving decoupling theorems or scaling
laws in the case of non-UV singularities is to elaborate the
contents of the operators and their anomalous dimensions that are
responsible for non-UV singularities. We hope this line of
investigations would lead to a different approach to the problem
of non-UV singularities\footnote{In QCD, this means an approach
for tackling the nonperturbative dynamics.} and help to clarify
the universal contents of factorization and/or its violation.

In summary, we derived the scaling laws in any EFT assuming the
existence of nontrivial underlying structures with renormalization
group equation and Callan-Symanzik equation being interpreted as
'decoupling' theorems of the underlying structures. The
Appelquist-Carazzone theorem was briefly discussed in this
underlying structures perspective.

\section*{acknowledgments}The author is grateful to Professor
Bernd A. Kniehl for his hospitality at the II. Institute for
Theoretical Physics of Hamburg University. This project is
supported in part by the National Natural Science Foundation of
China under Grant No.s 10205004 and 10475028, and by China
Scholarship Council.


\end{document}